\documentclass{appolb}
\usepackage{amsmath}
\usepackage{amssymb}
\usepackage{graphicx}
\usepackage{cite}

\begin{document}
\title{
Saturation in central-forward jet production\\ in p-Pb collisions at LHC
\thanks{Presented at the 42. International Symposium on Multiparticle Dynamics
(ISMD), 16-21 September 2012, Kielce, Poland.}
}
\author{
  Sebastian Sapeta 
  \address{Institute for Particle Physics Phenomenology, \\ Durham University,
           South Rd, Durham DH1 3LE, UK}
}
\maketitle
\vspace{-25em}
\begin{flushright}
  IPPP/12/97\\
  DCPT/12/194\\
\end{flushright}
\vspace{20em}

\begin{abstract}
We show that saturation can manifest itself in central-forward dijet production
in p-A collisions. In spite of large transverse momenta of the jets, the almost
back-to-back dijet configurations are able to probe gluon density at low $x$ and
low $k_t$. We perform our study in the framework of high energy factorization
with the unintegrated gluon density given by a nonlinear QCD evolution equation.
We show that the formalism can successfully account for features measured in e-p
and p-p data and it predicts significant suppression of the central-forward jet
decorrelations in p-Pb compared to p-p, which we attribute to saturation of
gluon density in the nucleus.
\end{abstract}

  
\section{Introduction}

Nonlinear effects of QCD dynamics are expected to become increasingly important
as one goes for a more and more forward production. This is because the forward
region corresponds to the density of the incoming gluon being probed at small
values of the longitudinal momentum fraction, $x$, for which a typical $k_t$ of
the gluon is of the order of, or smaller than, the saturation scale $Q_s(x)$. To
further enhance the relative importance of the saturation region, one may go to
the proton-nucleus (p-A) collision as the saturation scale in the nucleus is
expected to be significantly higher than in the proton.

The LHC experiments were designed predominantly to measure hard final states
like high-$p_t$ jets or vector bosons.
The framework that allows one to study dense gluon systems by the means of
looking at hard objects is provided by the high energy factorization formalism
\cite{Catani:1990eg}. In this contribution we show that saturation
effects may indeed be observed by looking at the azimuthal correlations
between two leading jets from p-A collisions, one of which is produced in the
forward direction.
Such a final state probes the parton density of the nucleus at low $x$, while
that of the proton at relatively large longitudinal momentum fraction.

\section{High energy factorization}

\begin{figure}[t] \centering
  \includegraphics[width=0.50\textwidth]{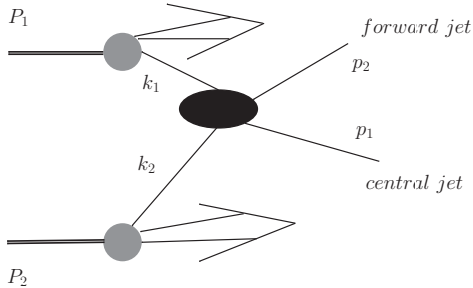}
  \caption{Jet production in the forward region in hadron-hadron collision.
  } 
  \label{fig:jet_production} 
\end{figure}

The leading order contribution to dijet production comes from the $2\to 2$
partonic process
\begin{equation}
  a (k_1) + b (k_2) \to c (p_1)+ d (p_2)\,.
\end{equation}
The fractions of the longitudinal momenta of the initial state partons are
related to the transverse momenta and rapidities of the final state partons by
$x_{1,2} = \frac{1}{\sqrt{S}} \left(p_{t1} e^{\pm y_1} + p_{t2} e^{\pm
y_2}\right)$ with $S$ being the squared energy of the incoming hadrons.
In the case of forward-central or forward-forward jet production, these
fractions of momenta are highly asymmetric, $x_1\simeq 1$ and $x_2 \ll 1$, we
can therefore neglect the transverse momentum of the parton~$a$. Using this fact
together with the Sudakov decomposition of 4-momenta of the incoming partons, we
arrive at the following formula for dijet production cross section in the
high energy factorization approach
\begin{multline} 
  \frac{d\sigma}{dy_1dy_2dp_{1t}dp_{2t}d\Delta\phi} 
  = \\
  \sum_{a,c,d} 
  \frac{p_{t1}p_{t2}}{8\pi^2 (x_1x_2 S)^2}
  {\cal M}_{ag\to cd}
  x_1 f_{a/A}(x_1,\mu^2)\,
  \phi_{g/B}(x_2,k^2)\frac{1}{1+\delta_{cd}}\,,
  \label{eq:cs-main}
\end{multline}
where $k^2 = p_{t1}^2 + p_{t2}^2 + 2p_{t1}p_{t2} \cos\Delta\phi$ and
$\Delta\phi$ is the azimuthal distance between the two outgoing partons.
All the notation in Eq.~(\ref{eq:cs-main}) corresponds to that in
Fig.~\ref{fig:jet_production}.  ${\cal M}_{ag\to cd}$ matrix element for the
$2\to 2$ process with one off-shell initial state gluon was calculated
in~\cite{Deak:2009xt}.
Since the parton $a$ is probed at high values of $x_1$, it is legitimate to
describe it with the collinear parton density $f_{a/A}(x_1,\mu^2)$. 
In our study we used one of the standard pdfs CTEQ6mE~\cite{Pumplin:2002vw}. 
On the other hand, for the incoming gluon, which is probed at small $x_2$, we
must use the unintegrated gluon density $\phi_{g/B}(x_2,k^2)$, which in addition
depends on that gluon's transverse momentum.
The sum in Eq.~(\ref{eq:cs-main}) runs over the following sub-processes: $qg
\to  qg$, $gg  \to  q\bar q$ and $gg \to gg$.

\section{Unintegrated gluon distribution in proton and nucleus}

In our study, we used the unintegrated gluon density from the unified BK/DGLAP
framework~\cite{Kwiecinski:1997ee, Kutak:2003bd,Kutak:2004ym}.  This approach
provides a well behaved, nonlinear gluon distribution inside the proton and 
it incorporates the main sources of higher order effects. 
We generalized this framework to the case of a nucleus by assuming the
Wood-Saxon nuclear density profile with the nucleus radius given by
$R_{\text{A}}=R\, A^{1/3}$ where $R$ is the proton radius and $A$ is the mass
number. In the limit $A\rightarrow 1$ the result for the proton is recovered. 
The corresponding evolution equation for the distribution of gluons per nucleon
in the nucleus $A$ reads
\begin{multline} 
\phi(x,k^2) \; = \; \phi^{(0)}(x,k^2) \\
+\,\frac{\alpha_s N_c}{\pi}\int_x^1\!\frac{dz}{z} \int_{k_0^2}^{\infty}\!
\frac{dl^2}{l^2}   \bigg\{\frac{l^2\phi(\frac{x}{z},l^2), \theta(\frac{k^2}{z}-l^2)  -
k^2\phi(\frac{x}{z},k^2)}{|l^2-k^2|}   +
\frac{k^2\phi(\frac{x}{z},k^2)}{|4l^4+k^4|^{\frac{1}{2}}}
\bigg\} \\  
+ \, \frac{\alpha_s}{2\pi k^2} \int_x^1\! dz \,\Bigg[ 
\left(P_{gg}(z)-\frac{2N_c}{z}\right) \int^{k^2}_{k_0^2} d l^{
2}\,\phi\left(\frac{x}{z},l^2\right)+zP_{gq}(z)\Sigma\left(\frac{x}{z},k^2\right)\Bigg] \\
-\frac{2A^{1/3}\alpha_s^2}{R^2}\left[\left(\int_{k^2}^{\infty}\frac{dl^2}{l^2}\phi(x,l^2)\right)^2
+
\phi(x,k^2)\int_{k^2}^{\infty}\frac{dl^2}{l^2}\ln\left(\frac{l^2}{k^2}\right)\phi(x,l^2)\right]\,,
\label{eq:fkovres2} 
\end{multline} 
where we used the simplified notation: $\phi(x,k^2) \equiv \phi_{g/A}(x,k^2)$
and $\alpha_s \equiv \alpha_s(k^2)$.
The second line in Eq.~(\ref{eq:fkovres2}) corresponds to the BFKL equation with
the kinematic constraint~\cite{Kwiecinski:1997ee} (introduced by the theta
function). The third line supplements DGLAP-type corrections and the last line
corresponds to the nonlinear term whose strength is enhanced for the nucleus by
the factor~$A^{1/3}$.

We started off by taking Eq.~(\ref{eq:fkovres2}), with an appropriate initial
condition, and fitted it to the most recent combined HERA
data~\cite{Aaron:2009aa}, which allowed us to fix the free parameters of the
framework (proton radius $R$ and three other coming from the parametrization of
the initial condition).  We performed the fit in the kinematical range of
$x<0.01$ and the full range of $Q^2$ and obtained a very good description of
data, with $\chi^2/\text{ndof} = 1.73$. For detailed discussion we refer to
\cite{Kutak:2012rf}.

\section{Signatures of saturation in dijet production in p-A collisions} 
\label{sec:results-p-Pb}

\begin{figure}[t]
  \centering
  \includegraphics[height=0.5\textwidth,angle=-90]{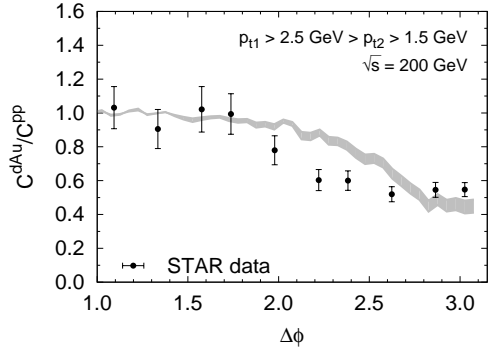}
  \caption{
  Ratio of d-Au/p-p coincidence probabilities $C(\Delta\phi)$ for the forward
  dihadron production at RHIC as a function of the azimuthal distance between
  the particles.  The band corresponds to our prediction with the uncertainty
  related to unknown yield of uncorrelated dihadron production.  For full
  details see~\cite{Kutak:2012rf}.
}
  \label{fig:decor-d-Au}
\end{figure}

Once the framework for the unintegrated gluon density had been fully specified,
we used it to study saturation effects in the dijet production in p-A
collisions.
One observable which is very well suited for studying these effects is the
distribution of dijet azimuthal distance $\Delta\phi$. In the region $\Delta\phi
\sim \pi$, which corresponds to the two jets being produced almost back-to-back,
the gluon density is probed at low $k_t$ where the nonlinear effects are strong
and the gluon distribution is expected to be significantly suppressed.

Even though our main focus was on the central-forward dijet production at the
LHC, the framework described above allowed us also to make an estimate of the
suppression of dihadron production in d-Au collisions at RHIC.
Fig.~\ref{fig:decor-d-Au} shows the ratio of the normalized $\pi^0$ yields in
the d-Au and the p-p collisions from STAR~\cite{Braidot:2010zh}. There, we also
show our prediction which follows directly from using Eq.~(\ref{eq:cs-main})
together with the unintegrated gluon distribution from Eq.~(\ref{eq:fkovres2})
with all the parameters fixed by HERA data and the mass number set to A=196 (for
detailed procedure see \cite{Kutak:2012rf}).
As we see, our prediction correctly reproduces the suppression observed
at RHIC in the region $\Delta\phi \sim \pi$, which shows that our theoretical
framework captures the essential physics of this class of processes.

\begin{figure}[t]
  \centering
  \includegraphics[width=0.42\textwidth,angle=-90]{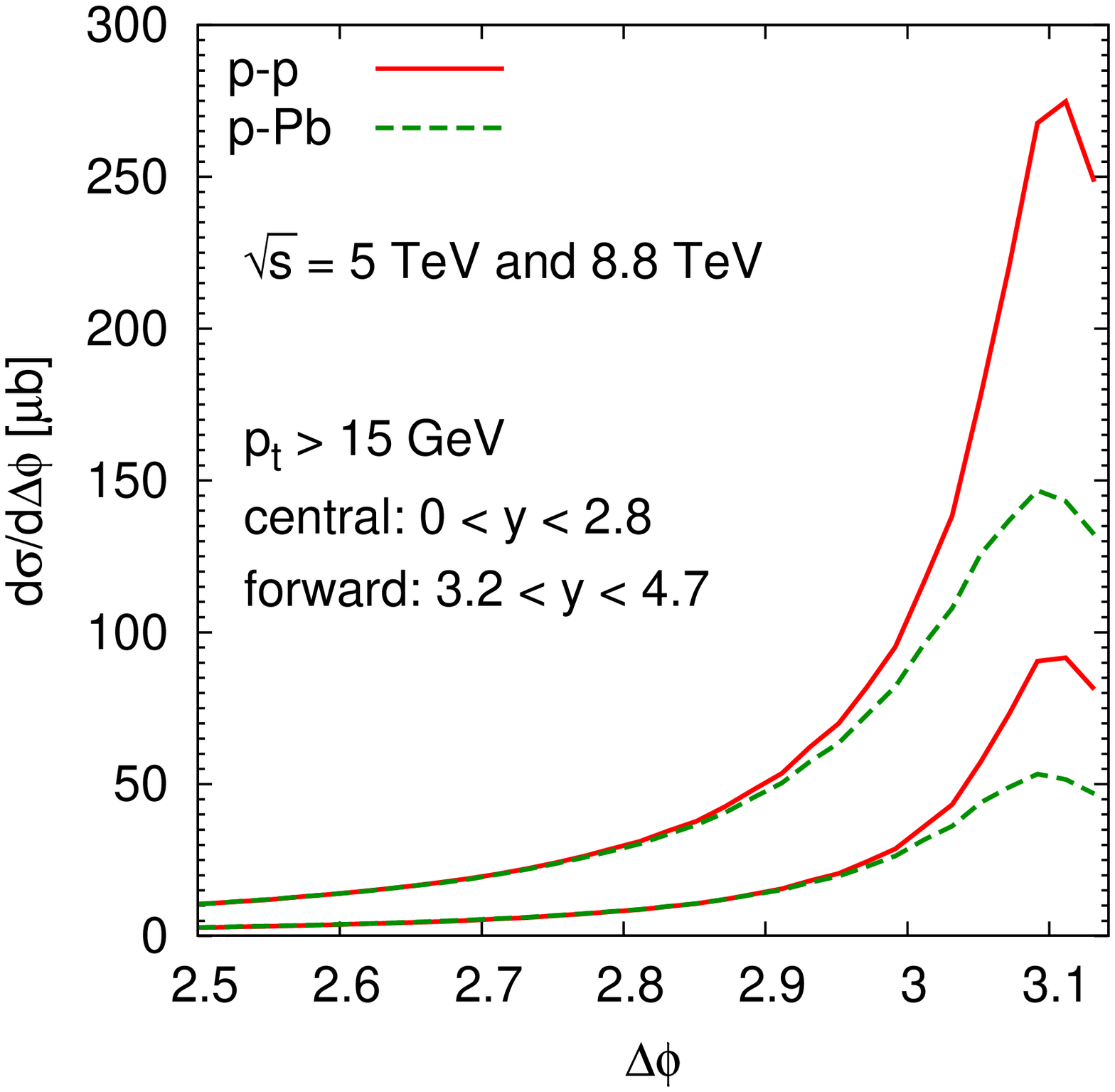}
  \hskip 20pt
  \includegraphics[width=0.42\textwidth,angle=-90]{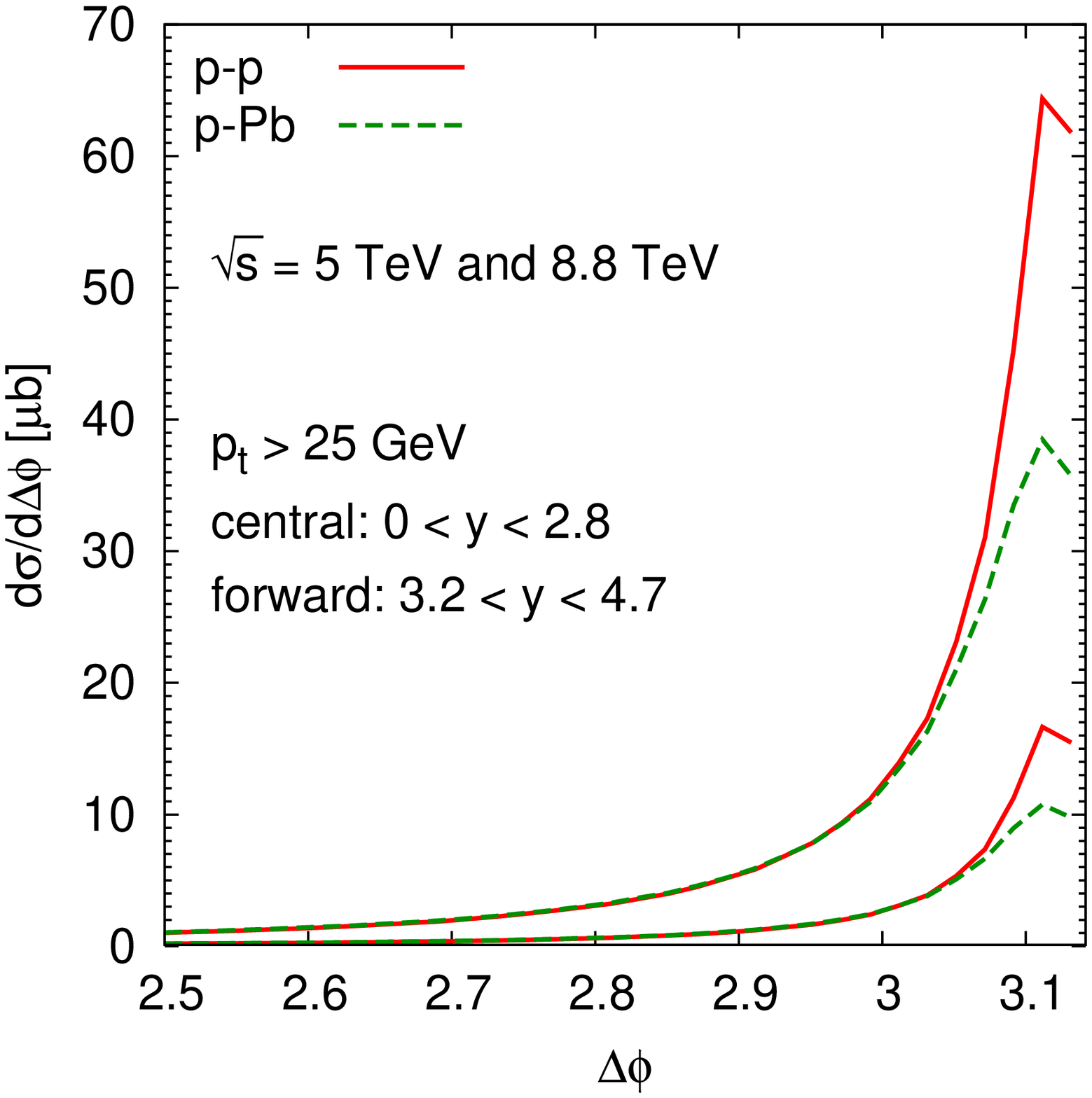}
  \caption{ 
  Differential cross sections for central-forward dijet production at $\sqrt{s}
= 5 \text{ TeV}$ and 8.8~TeV as functions of azimuthal distance between the jets
$\Delta\phi$ for the case of p-p and p-Pb collisions and two different cuts on
jets' $p_t$.
}
  \label{fig:decor-p-Pb-588}
\end{figure}

We then moved to the central-forward dijet production in the p-Pb collisions at
the LHC. In Fig.~\ref{fig:decor-p-Pb-588} we show the corresponding cross
section as a function of the azimuthal distance between the jets for two
energies of the p-Pb collisions, i.e. the current $\sqrt{s} = 5\, \text{TeV}$
and the nominal $\sqrt{s} = 8.8\, \text{TeV}$. We also used two different jet
$p_t$  cuts 15 and 25 GeV.

We see that the $\Delta\phi$ distribution in the peak region is suppressed by
the factor two for the case of the p-Pb collision with respect to p-p,
and the effect extends to lower values of $\Delta\phi$ as we lower the jet $p_t$
threshold. 
This is a consequence of gluon saturation which is stronger in the Pb nucleus
then in the proton and leads to suppression of gluon density at low $k_t$, which
is the region probed by the dijet configurations with $\Delta\phi \sim \pi$ .
As expected, the total yields increase with energy and decrease with $p_t$ cut
but the relative difference between the p-p and p-Pb case remains similar.

The distributions shown in Fig.~\ref{fig:decor-p-Pb-588} could be subjet to
further corrections coming from the Sudakov and parton shower effects.
They are however expected to act in a similar way for the
proton and for the heavy ion collision since they affect the hard scattering. 
We would like to emphasize that the suppression observed in
Fig.~\ref{fig:decor-p-Pb-588} comes from the initial state parton density
saturation and therefore the relative difference between the cases of p-p and
p-Pb will persist even if the very small region near $\Delta\phi=\pi$
may profit from further refinements.

\section{Conclusions}

We presented predictions for the azimuthal dijet decorrelations in the p-p and
p-A collisions in the framework of the high energy factorization with the
unintegrated gluon density given by the nonlinear QCD evolution equation. This
approach is unique as it allows one to study hard final states in the framework
with saturation.

We validated our approach by using HERA $F_2$ data and we checked that it
correctly estimates the suppression of the away side peak observed by STAR.
For the LHC, we found that the saturation in the Pb nucleus has a potential to
manifest itself as a factor two suppression of the central-forward jet
decorrelation in the region of the azimuthal distance between the jets
$\Delta\phi \sim \pi$. We argued that this relative difference should be largely
insensitive to the final state effects.

\section*{Acknowledgments}

We thank the organizers of the 42. ISMD conference in Kielce for the very
interesting and stimulating meeting.
The original results presented here were obtained with Krzysztof
Kutak and partially supported by the Foundation for Polish Science with
the grant Homing Plus/2010-2/6.

\end{document}